\documentclass[12pt]{article}

\usepackage{amsmath, cite}

\begin{document}

\title{Some results on finite amplitude elastic waves propagating in 
rotating media}
\author{Michel Destrade, Giuseppe Saccomandi}
\date{2004}
\maketitle

\bigskip

\begin{abstract}

Two questions related to elastic motions are raised and addressed. 
First: in which theoretical framework can the equations of motion 
be written for an elastic half-space put into 
uniform rotation?
It is seen that nonlinear finite elasticity provides such a framework 
for incompressible solids.
Second: how can finite amplitude exact solutions be generated? 
It is seen that for some finite amplitude transverse waves in
rotating incompressible elastic solids with general shear
response, the solutions are obtained by reduction of the equations of 
motion to a system of ordinary differential equations
equivalent to the system governing the central motion problem of 
classical mechanics. 
In the special case of circularly-polarized harmonic progressive 
waves, the dispersion equation is solved in closed form for a variety 
of shear responses, 
including nonlinear models for rubberlike and soft biological tissues. 
A fruitful analogy with the motion of a nonlinear string is pointed 
out.

\end{abstract}

\newpage

\section{Introduction}

The propagation of elastic waves in rotating media has been a subject 
of continuous interest in the last three decades or so. 
Ever since the publication of a seminal article by Schoenberg and 
Censor \cite{ScCe73},
numerous workers have studied how uniform rotation affects 
time-dependent solutions to the governing equations (pointers to such 
studies can be found in recent articles on waves in rotating media 
such as Refs.\cite{AhKh01, ZhJi01,Auri04,Dest04}.) 
The starting point of these studies is the inclusion
of the Coriolis and centrifugal accelerations into the equations of 
motion:
\begin{equation}  \label{1.1}
\text{div } \mathbf{T} = \rho \ddot{\mathbf{y}} + 2 \rho
\mbox{\boldmath
$\Omega$} \times \dot{\mathbf{y}} + \rho \mbox{\boldmath $\Omega$} 
 \times (\mbox{\boldmath $\Omega$} \times \mathbf{y}).
\end{equation}
Here $\mathbf{T}$ is the Cauchy stress tensor, 
$\rho$ is the mass density, 
$\mathbf{y} = \mathbf{y}(\mathbf{x},t)$ denotes the current position 
of a particle in the material initially at $\mathbf{x}$ in the 
reference configuration, and $\mbox{\boldmath $\Omega$}$ is the constant rotation 
rate vector.
Also,  a dot denotes differentiation with respect to 
time $t$ in a \textit{fixed} (non-rotating) frame; 
in other words, if ($\mathbf{e_1}$, $\mathbf{e_2}$, $\mathbf{e_3}$)
is one such frame, then $\mathbf{y} = y_i \mathbf{e_i}$ and 
$\dot{\mathbf{y}} := (\partial y_i/\partial t)\mathbf{e_i}$.

The second term on the right hand-side of Eq.~\eqref{1.1} is the 
Coriolis force and the third term is the centrifugal force. 
This latter term is the source of an obvious concern in a linearly 
elastic material with infinite dimension(s) because it grows linearly 
with the distance between the particle and the axis of rotation. 
Most (and perhaps all) previous works on the subject have dealt with 
this potential problem simply by focusing on the so-called 
``time-dependent'' part of the equations of motion. 
In this approach, the solution $\mathbf{y}$ is split into a 
``time-independent'' part and a ``time-dependent'' part as 
$\mathbf{y}(\mathbf{x},t) 
 = \mathbf{y^s}(\mathbf{x}) + \mathbf{u}(\mathbf{x},t)$ (say). 
Then, the constitutive equation of the elastic material being linear, 
the Cauchy stress can also be split: 
$\mathbf{T}(\mathbf{x},t) = \mathbf{T^s}(\mathbf{x}) 
+ \mbox{\boldmath $\sigma$}(\mathbf{x},t)$ (say) and the linearization 
of the equations of motion allows for the \textit{separate} resolution 
of a time-independent problem and of a time-dependent problem,
\begin{align} 
& \text{div } \mathbf{T^s}(\mathbf{x}) 
= \rho \mbox{\boldmath $\Omega$}
 \times (\mbox{\boldmath $\Omega$} \times \mathbf{y^s}(\mathbf{x})),  
 \label{1.2}
\\
& \text{div } \mbox{\boldmath $\sigma$}(\mathbf{x},t) 
= \rho \ddot{\mathbf{u}}(\mathbf{x},t) 
 +2\rho \mbox{\boldmath $\Omega$} \times \dot{\mathbf{u}}(\mathbf{x},t)
 + \rho \mbox{\boldmath $\Omega$} 
     \times (\mbox{\boldmath$\Omega$}\times \mathbf{u}(\mathbf{x},t)).
 \label{1.3}
\end{align}
Although the resolution of Eq.~\eqref{1.3} has generated a
wealth of results in a variety of contexts, the resolution of 
Eq.~\eqref{1.2} seems to have been left aside, at least as long as 
potentially infinite distances from the rotation axis are involved. 
This paper aims at providing a context in which the \textit{global}
equations of motion in a rotating elastic media Eq.~\eqref{1.1}, 
possibly inclusive of finite strain effects and of a nonlinear 
constitutive equation, can be posed and solved.

Because large strains might appear in a rotating elastic solid, we 
place ourselves in the framework of finite nonlinear elasticity. 
We focus on materials subject to the internal constraint of 
incompressibility, first because many actual materials with a 
nonlinear elastic response such as rubber or biological soft tissue 
can be considered to be incompressible, and second because the 
inherent introduction of an arbitrary scalar quantity 
(the ``pressure'') leads to an immediate simplification of the 
equations of motion Eq.~\eqref{1.1}. 
Indeed, as we show in the next Section, the arbitrariness of the 
$p\mathbf{1}$ term in the constitutive equation of an
incompressible body allows for the centrifugal force to be absorbed by 
this pressure term. 
Once this manipulation is done, the resolution of the
equations of motion can be conducted quite naturally. 
As noted by Schoenberg and Censor \cite{ScCe73}, two features 
characterize waves in rotating bodies as opposed to waves in 
non-rotating bodies: a new direction of anisotropy
(linked to the rotation axis) and more dispersion (linked to the 
rotation frequency). 
To illustrate these features, we revisit some classic results on
finite amplitude elastic motion due to Carroll \cite{C1,C1a,C2,C3} and
extend them to the case of a body in rotation.

The exact solutions of Carroll are versatile in their fields of 
application because they are valid not only for nonlinearly 
elastic solids, but also for viscoelastic solids
\cite{C4}, Reiner-Rivlin fluids \cite{C4, C5}, Stokesian fluids 
\cite{C4}, Rivlin-Ericksen fluids \cite{C5}, liquid crystals 
\cite{C6}, dielectrics \cite{C7}, magnetic materials \cite{C8}, etc. 
They also come in a great variety of forms,  
as circularly-polarized harmonic progressive waves, as motions with 
sinusoidal time dependence, as motions with sinusoidal space 
dependence, etc.
In our revisiting his findings, we note
a striking analogy between the equations of motion obtained for a 
motion general enough to include all of the above motions, and the 
equations obtained in the problem of the motion of a
nonlinear string, as considered by Rosenau and Rubin \cite{RR}. 
Then we show how the method of \cite{RR} can be used to derive all 
(and more) of the different results obtained by Carroll, which turn out 
to be a direct consequence of material isotropy and of the Galilean 
invariance of the field equations.

The paper is organized in the following manner. 
In the next Section the basic equations for motions in a rotating 
nonlinearly elastic incompressible solid 
and their specialization to finite amplitude transverse waves are 
given.
In Section 3 we recast the determining
equations in a general complex form and we show that they admit
some special separable solutions. 
In Section 4 we investigate in detail the case of
circularly-polarized harmonic progressive waves. 
We give the dispersion relation and solve it 
for Mooney-Rivlin materials and for some other strain energy density 
functions relevant to the modelling of rubberlike materials (some of 
these results are new even in the non-rotating case). 
Next we show that in rotating solids, motions with a sinusoidal time 
dependence (Section 5) and motions with a sinusoidal spatial 
dependence (Section 6) are determined by solving a reduced system of 
ordinary differential equations, equivalent to that of a central 
motion problem. 
The main difference with Carroll's results for the non-rotating case 
is that, for special values of the angular velocity, the central force 
may be repulsive; 
this possibility is ruled out in the non-rotating case by the
empirical inequalities \cite{BB}.

\section{Preliminaries}

\subsection{Equations of motion in a rotating elastic solid}

Let the initial and current coordinates of a point of the body, 
referred to the same fixed rectangular Cartesian system of axes, 
be denoted by $x_{i}$ and $y_{i}$, respectively, where the indices 
take the values $1$, $2$, $3$.
A motion of the body is defined by
\begin{equation}  \label{1}
\mathbf{y}=\mathbf{y}(\mathbf{x},t).
\end{equation}
The response of a homogeneous isotropic incompressible elastic solid to
deformations from an undistorted reference configuration is described 
by the constitutive relation,
\begin{equation}  \label{2}
\mathbf{T} = -\widetilde{p} \mathbf{1} 
 + \alpha \mathbf{B}-\beta \mathbf{B}^{-1},
\end{equation}
where $\mathbf{T}$ is the Cauchy stress tensor, $\mathbf{1}$ is the 
unit tensor, and $\mathbf{B}$ is the left Cauchy-Green strain tensor, 
defined by
\begin{equation}  \label{3}
\mathbf{B} := \mathbf{FF}^{T},
\end{equation}
$\mathbf{F} := \partial \mathbf{y}/\partial \mathbf{x}$ being the 
deformation gradient tensor. 
Also in Eq.~\eqref{2}, $\widetilde{p}$ is an arbitrary scalar
function associated with the internal constraint of incompressibility
\begin{equation}  \label{4}
\det \mathbf{F} = 1,
\end{equation}
to be determined from the equations of motion and eventual 
boundary/initial conditions. 
The response parameters $\alpha $ and $\beta $ are functions of
the first and second invariants of $\mathbf{B}$: 
$\alpha = \alpha(I,II)$, $\beta = \beta(I,II)$, where
\begin{equation}  \label{5}
I=\text{tr }\mathbf{B}, \quad II=\text{tr }\mathbf{B}^{-1}.
\end{equation}
For a hyperelastic material, a strain energy density per unit of 
volume $W=W(I,II)$ is defined and $\alpha$, $\beta$ are given by
\begin{equation}  \label{6}
\alpha = 2\frac{\partial W}{\partial I}, \quad 
\beta = 2\frac{\partial W}{\partial II}.
\end{equation}

Now we consider that the elastic medium rotates with a uniform 
rotation vector $\mathbf{\Omega }$, about a given axis. 
In the absence of body forces, the equations of motions relative to a 
rotating frame (see for instance \cite[pp.60--61]{Liu02}) are given by 
Eq.\eqref{1.1}. 
Using the constitutive equation Eq.~\eqref{2}, we obtain
\begin{equation}  \label{7}
-\text{grad } \widetilde{p} 
 + \text{div } (\alpha \mathbf{B} - \beta \mathbf{B}^{-1}) 
= \rho \ddot{\mathbf{y}} 
  + 2 \rho \mbox{\boldmath $\Omega$} \times \dot{\mathbf{y}}
   + \rho \mbox{\boldmath $\Omega$} \times (\mbox{\boldmath$\Omega$} 
                                                  \times \mathbf{y}).
\end{equation}
Now write $\widetilde{p}$ in the form
\begin{equation}  \label{9}
\widetilde{p} = p 
 - \textstyle{\frac{1}{2}}\rho [\mathbf{\Omega \times} 
                  (\mathbf{\Omega\times y})] \mathbf{\cdot y},
\end{equation}
where $p = p(\mathbf{x},t)$ is yet another arbitrary pressure scalar. 
Then Eq.~\eqref{7} reduces to
\begin{equation}  \label{10}
-\text{grad } p
   + \text{div } (\alpha \mathbf{B} - \beta \mathbf{B}^{-1}) 
  = \rho \ddot{\mathbf{y}} 
  + 2 \rho \mbox{\boldmath $\Omega$} \times \dot{\mathbf{y}}.
\end{equation}
Hence the equations of motion can be tackled independently of the
centrifugal acceleration, which does not appear here. 
Once Eqs.\eqref{10} are solved, the solution $\mathbf{y}$ will lead to 
a pressure field $\widetilde{p}$ given by Eq.~\eqref{9}
which does depend on the centrifugal force.

\subsection{Finite amplitude shearing motions}

Following Carroll \cite{C1}, we study for the remainder of the paper 
the propagation of plane transverse waves in a bi-axially deformed
incompressible material. 
Thus we consider the following class of shearing motions,
\begin{equation}
y_1 = \mu x_1 + u(z,t), \quad 
y_2 = \mu x_2 + v(z,t), \quad 
y_3 = \lambda x_3 =:z,  \label{11}
\end{equation}
that is, a transverse wave polarized in the ($x_1 x_2$) plane and
propagating in the $x_3$-direction of a material subject to a pure
homogeneous pre-stretch with constant principal stretch ratios $\mu $, 
$\mu$, $\lambda$ ($\mu^2 \lambda =1$) in the $x_{1}$, $x_{2}$, $x_{3}$
directions, respectively. 
For these motions, we find
\begin{equation}
\mathbf{B}=
\begin{bmatrix}
\mu^2 + \lambda^2 u_z^2 &  &  \\
\lambda^2 u_z v_z & \mu^2 + \lambda^2 v_z  &  \\
\lambda^2 u_z  & \lambda^2 v_z  & \lambda^2
\end{bmatrix}, 
 \quad 
\mathbf{B}^{-1}=
  \begin{bmatrix}
    \lambda &  &  \\
     0 & \lambda &  \\
    -\lambda u_z & -\lambda v_z & \lambda (u_z^2 + v_z^2)+\mu^4
  \end{bmatrix}.
\end{equation}
Here and henceforward, a subscript letter denotes partial 
differentiation (i.e. $u_{z}:=\partial u/\partial z$, 
$v_{tt}:=\partial ^{2}v/\partial t^{2}$, etc.) 
It follows from Eq.~\eqref{5} that
\begin{equation}  \label{12}
I = 2 \mu^2 + \lambda^2 (1+u_z^2 + v_z^2), \quad 
II = \mu^4 + \lambda (2 + u_z^2 + v_z^2),
\end{equation}
so that both invariants, and consequently the response parameters 
$\alpha$, $\beta$, are functions of $u_z^2 + v_z^2$ alone,
\begin{equation}
\alpha = \alpha (u_z^2 + v_z^2),\quad 
\beta = \beta (u_z^2 + v_z^2).
\end{equation}
Then the equations of motion Eq.~\eqref{10} read
\begin{align}
& -p_{y_1} + (Qu_z)_z = \rho (u_{tt} - 2\Omega_3 v_t),  
\notag \label{13} \\
& -p_{y_2} + (Qv_z)_z = \rho (v_{tt} + 2\Omega _3 u_t),  
\notag \\
& -p_z + [\alpha \lambda^2 + \beta \mu^4 
   + \beta \lambda (u_z^2 + v_z^2)]_z 
 = 2\rho (\Omega_1 v_t - \Omega_2 u_t),
\end{align}
where the function $Q=Q(u_z^2 + v_z^2)$ is defined by
\begin{equation}  \label{15}
Q:= \alpha \lambda^2 + \beta \lambda.
\end{equation}
By inspection of Eqs.~\eqref{13}, we find that $p$ can be taken in the 
form
\begin{equation}
p = p(z,t) = 
  \alpha \lambda^2 + \beta \mu^4 - \beta \lambda (u_z^2 + v_z^2)
   - 2 \rho \textstyle{\int }(\Omega_1 v_t - \Omega_2 u_t) \text{d}z.
\label{16}
\end{equation}
Then Eq.~\eqref{13}$_{3}$ is satisfied and Eqs.~\eqref{13}$_{1,2}$ 
reduce to
\begin{equation}  \label{motion}
(Qu_z)_z = \rho (u_{tt} - 2\Omega_3 v_t),
 \quad 
(Qv_z)_z = \rho (v_{tt} + 2\Omega_3 u_t).
\end{equation}

Eqs.~\eqref{motion} form a system of two coupled nonlinear hyperbolic
partial differential equations, generalizing the system derived by 
Carroll in \cite{C1} for a non-rotating body.

\section{Separable solutions}

\subsection{Link with another problem (string motion)}

By inspection of the system Eqs.~\eqref{motion}, an analogy can be 
drawn with the system of equations governing the motion of a nonlinear 
string, as treated by Rosenau and Rubin \cite{RR}. 
Indeed, if the position of a particle in a string is denoted by the 
rectangular Cartesian coordinates $x(\xi ,t)$, $y(\xi ,t)$, where 
$\xi$ is a curvilinear abscissa, then the equations of
motion of the string can be put in the form,
\begin{equation}  \label{string}
\left[ (T/a) x_\xi \right]_\xi = \rho_0 (x_{tt} - f_1), \quad 
\left[ (T/a) y_\xi \right]_\xi = \rho_0 (y_{tt} - f_2).
\end{equation}
Here, $T$ is the internal tension in the string (acting along the 
tangent to the string curve), $f_{1}$ and $f_{2}$ are the components 
of the body force per unit mass, $\rho_0 = \rho_0(\xi)$ is the mass 
density, and $a$ is the metric associated with the stretch of the 
string: $a=\sqrt{ x_\xi^2 + y_\xi^2}$. 
Finally, a constitutive equation $T=T(a)$ for the internal tension
characterizes a the string material.

The similarity between the two systems Eqs.~\eqref{motion} and 
Eqs.~\eqref{string} is striking. 
Accordingly we now adapt the analysis devised by Rosenau and Rubin
\cite{RR} for a nonlinear string to our system of governing equations.

\subsection{Separation of variables}

Seeking some exact solutions, we follow Rosenau's and Rubin's \cite{RR}
steps. 
First we differentiate Eqs.~\eqref{motion} with respect to $z$, 
and obtain
\begin{equation}  \label{motion1}
[ Q U]_{zz} = \rho (U_{tt} - 2\Omega_3 V_t), \quad 
[Q V]_{zz} = \rho (V_{tt} + 2\Omega_3 U_t),
\end{equation}
where $U := u_z$ and $V := v_z$. 
Next, we define the complex function $Z$ as
\begin{equation}  \label{ses}
Z(z,t) =\eta (z,t) \text{e}^{i\xi (z,t)} := U + iV,
\end{equation}
so that
\begin{equation}
U = \Re(Z) = \eta \cos \xi, \quad V = \Im(Z) = \eta \sin \xi.  
\label{ses1}
\end{equation}
Then, we rewrite the system Eqs.~\eqref{motion1} as a single complex 
equation,
\begin{equation} 
[Q (\eta^2) Z]_{zz} = \rho (Z_{tt} + 2i \Omega_3 Z_t).  
\label{ses2}
\end{equation}
To reduce further this equation to a set of ordinary differential 
equations, we look for a class of solutions admitting the separable 
forms:
\begin{equation}  \label{ses3}
\eta(z, t) = \eta_1(z) \eta_2 (t), \quad 
\xi(z,t) = \xi_1(z) + \xi_2 (t),
\end{equation}
where $\eta_1$ and $\xi_1$ ($\eta_2$ and $\xi_2$) are functions of 
space (time) only. 
Then Eq.~\eqref{ses2} can be cast in the form
\begin{equation}  \label{ses4}
\dfrac{[Q(\eta_1^2 \eta_2^2) \eta_1 \text{e}^{i\xi_1}]_{zz}} 
          {\eta_1 \text{e}^{i \xi_1}} 
 = \rho \dfrac{(\eta_2 \text{e}^{i\xi_2})'' +
    2i\Omega_3 (\eta_2 \text{e}^{i \xi_2})'}
                    {\eta_2 \text{e}^{i\xi_2}},
\end{equation}
where the prime denotes differentiation with respect to the argument 
of a single-variable function.

Rosenau and Rubin \cite{RR} noted that a sufficient condition to ensure
complete separation of time functions from space functions in this 
equation is that the material response function $Q$ be itself 
separable. 
Indeed if
\begin{equation}  \label{ses5}
Q(\eta_1^2 \eta_2^2) = Q_1(\eta_1^2) Q_2(\eta_2^2),
\end{equation}
(say) then we end up with the two ordinary differential equations,
\begin{align}  \label{Q1Q2}
& [Q_1 (\eta_1^2) \eta_1 \text{e}^{i\xi_1}]''
    = h \eta_1 \text{e}^{i\xi_1},  
\notag \\
& \rho [(\eta_2 \text{e}^{i\xi_2})'' 
   + 2i \Omega_3 (\eta_2 \text{e}^{i\xi_2})'] 
  = h Q_2(\eta_2^2) \eta_2 \text{e}^{ i\xi_2},
\end{align}
for some constant $h$.

The separation condition Eq.~\eqref{ses5} is however rather strong and 
might be fulfilled only for very specific constitutive equations. 
Another possibility, not mentioned by Rosenau and Rubin, 
for the separation of space functions from time functions  
arises when either $\eta_1(z)$ or $\eta_2(t)$ are constant functions 
(independent of their argument). 
Hence, when $\eta_1 = k_1$ (say), Eq.~\eqref{ses4} yields
\begin{equation}  \label{eta1Const}
( \text{e}^{i\xi_1})'' = h \text{e}^{i\xi_1}, \quad 
\rho [(\eta_2 \text{e}^{i\xi_2})'' 
   + 2i \Omega_3 (\eta_2 \text{e}^{i\xi_2})'] 
     = h Q (k_1^2\eta_2^2) \eta_2 \text{e}^{i\xi_2},
\end{equation}
and when $\eta_2 = k_2$ (say), it yields
\begin{equation}  \label{eta2Const}
[Q (k_2^2\eta_1^2) \eta_1 \text{e}^{i\xi_1}]'' 
     = h \eta_1 \text{e}^{i\xi_1}, 
\quad 
\rho [(\text{e}^{i\xi_2})'' + 2i \Omega_3 (\text{e}^{i\xi_2})'] 
      = h \text{e}^{ i\xi_2}.
\end{equation}
The conditions $\eta_1=$const. or $\eta_2 =$const. do not impose any
restriction on the strain energy function. 
Thus, the solutions to Eqs.~\eqref{eta1Const} or Eqs.~\eqref{eta2Const} 
are valid for any type of material,
in contrast to the solutions to Eqs.~\eqref{Q1Q2}, which require 
Eq.~\eqref{ses5} to be satisfied.

For instance, consider the solution
\begin{equation}
Z(z,t) = [\psi(t) + i \phi(t)] k \text{e}^{i(kz + \theta(t))}, 
 \label{ses13}
\end{equation}
where $k$ is a constant and $\psi$, $\phi$, $\theta$ are arbitrary real
functions of time. 
A simple check shows that $Z$ is indeed of the form given by 
Eqs.~\eqref{ses} and Eqs.~\eqref{ses3}, with the following 
identifications:
$\eta_1(z) = k = $ const., 
$\eta_2(t) = [\phi^2 + \psi^2]^{\textstyle{\frac{1}{2}}}$, 
$\xi_1(z) = kz$, and $\xi_2(t) = \theta + \tan^{-1}(\phi/\psi)$.
Once the ordinary differential equations Eqs.~\eqref{eta1Const} are 
solved, the displacement field is given by
\begin{align}  \label{field1}
& u(z,t) = \phi(t) \cos (kz+\theta (t)) + \psi(t) \sin(kz+\theta (t)),
\notag \\
& v(z,t) = \phi (t)\sin (kz+\theta (t)) - \psi(t) \cos (kz+\theta (t)).
\end{align}

On the other hand, consider the solution
\begin{equation}
Z(z,t) = 
 [(i\phi(z) + \psi(z)) \theta'(z) + (\phi'(z) - i\psi'(z))]
                            \text{e}^{i(\omega t+\theta(z))},  
\label{ses14}
\end{equation}
where $k$ is a constant and $\psi $, $\phi $, $\theta $ are arbitrary
functions of space. 
Here $Z$ is of the form given by Eqs.~\eqref{ses} and 
Eqs.~\eqref{ses3}, with the identifications 
$\eta_1(z) = 
 [(\phi' + \psi \theta')^2 
      + (\phi \theta' - \psi')^2]^{\textstyle{\frac{1}{2}}}$, 
$\eta_2(t) = 1 =$ const., 
$\xi_1(z) = \theta + \tan^{-1}[(\phi \theta' - \psi')/
               (\psi \theta' + \phi')]$, 
and $\xi_2(t) = \omega t$. 
Once the ordinary differential equations Eqs.~\eqref{eta2Const}
are solved, the displacement field is given by
\begin{align}
& u(z,t) = \phi(z) \cos(\omega t + \theta(z))
              + \psi(z)\sin(\omega t + \theta(z)), 
 \notag  \label{field2} \\
& v(z,t) = \phi(z) \sin(\omega t + \theta(z))
              - \psi(z) \cos(\omega t + \theta(z)).
\end{align}

The two sets of displacement fields Eqs.~\eqref{field1} and 
Eqs.~\eqref{field2} provide a great variety of possible finite 
amplitude motions, valid in every deformed incompressible nonlinearly 
elastic solid. 
They are inclusive of the solutions discovered and analyzed by Carroll 
over the years. 
Thus the motion Eqs.~\eqref{field1} written at $\psi (t)=0$ 
corresponds to the ``oscillatory shearing motions'' treated in 
\cite{C1a}; 
the motion Eqs.~\eqref{field2} written
at $\psi (z)=0$ corresponds to the ``motions with time-independent
invariants'' treated in \cite{C1a}; 
the motion Eqs.~\eqref{field2} written at $\theta (z)=0$ corresponds 
to the ``motions with sinusoidal time dependence''
or ``finite amplitude circularly-polarized standing waves'' treated in 
\cite{C2,C3}; 
the motion Eqs.~\eqref{field1} written at $\phi (z)=$const., 
$\psi (z)=0$, $\theta (z)=-kz$, 
or equivalently the motion Eqs.~\eqref{field2} written at 
$\phi (t)=$const., $\psi (t)=0$, $\theta (t)=-\omega t$, corresponds 
to the celebrated finite-amplitude circularly-polarized harmonic 
progressive waves of \cite{C1}.

Before we consider in turn each of these finite-amplitude motions for a
rotating body, we sum up the main results established in this 
Section.
We used a formalism proposed by Rosenau and Rubin \cite{RR} for the 
plane motion of a nonlinear string to derive separable solutions to 
the equations of motion of a deformed rotating solid in which 
finite-amplitude shearing motions might propagate. 
In the process, we noticed that two classes of
solutions, not considered by Rosenau and Rubin, were valid for any 
form of the strain energy function. 
Each class provided solutions which generalize
those proposed by Carroll \cite{C1,C1a,C2,C3,C4,C5,C6,C7,C8} and which 
put them into a wider context. 
On the other hand, the complex formalism makes it clear 
that the solutions considered here are related to natural symmetry
properties of the coupled wave equations Eqs.~\ref{motion1}. 
These properties are natural because they come out from material 
symmetries and frame indifference requirements \cite{BB}. 
We refer to the works of Olver \cite{O} and of Vassiliou \cite{V} 
for further information on the application of group analysis to 
coupled wave equations.

\section{Circularly-polarized harmonic waves}

First we consider a finite amplitude circularly-polarized harmonic
progressive wave propagating in the $z$-direction,
\begin{equation}  \label{circle}
u(z,t) = A \cos (kz-\omega t), \quad v(z,t) = \pm A \sin (kz-\omega t),
\end{equation}
which is a subcase of Eqs.~\eqref{field1} or of Eqs.~\eqref{field2}. 
Here the amplitude $A$, the wave number $k$, and the frequency 
$\omega$ are real positive constants, 
and the plus (minus) sign for $v(z,t)$ corresponds to a
left (right) circularly-polarized wave. 
For the choice of motion Eqs.~\eqref{circle}, we have
\begin{equation}
u_z^2 + v_z^2 = A^2 k^2,
\end{equation}
and Eqs.~\eqref{motion} reduce to the following dispersion
equation,
\begin{equation}  \label{dispersion}
k^2 Q (A^2 k^2) = \rho (\omega^2 \mp 2\Omega_3 \omega).
\end{equation}
The actual explicit form of the dispersion depends on a given 
constitutive equation. 
However we recall that, according to considerations by Carroll
\cite{C1} pertaining to the non-rotating case, $k^2 Q (A^2 k^2)$ must 
be a positive, monotonically increasing function tending to infinity 
with $k^2$.
It follows from the dispersion equation Eq.~\eqref{dispersion}, that 
for a given left circularly-polarized wave, the rotation rate 
$\Omega_3$ has a 
cut-off frequency of $\omega /2$ and the wave does not exist for 
rotation rates $\Omega_3$ beyond that cut-off frequency.

We now treat in turn three types of constitutive equations, which have
proved useful for the modelling of some incompressible rubberlike and 
soft biological materials.

\subsection{Waves in deformed Mooney-Rivlin materials}

As a first illustration we consider a Mooney-Rivlin hyperelastic 
material, with strain energy density,
\begin{equation}  \label{MR}
W_\text{MR} = C(I-3)/2 + D(II-3)/2,
\end{equation}
where $C$ and $D$ are constants, satisfying \cite{BoHa95} 
$C>0$, $D\geq 0$ or $C \geq 0$, $D > 0$.
It follows at once from Eqs.~\eqref{6} that $\alpha =C$, $\beta =D$, 
and by Eq.~\eqref{15}, that $Q$ is also independent of $z$. 
Introducing the speed $c$ of circularly-polarized waves 
in a bi-axially deformed, non-rotating Mooney-Rivlin
material \cite{C1,BoHa95},
\begin{equation}
\rho c^2 := Q = C\lambda^2 + D\lambda,
\end{equation}
we find that the dispersion equation Eq.~\eqref{dispersion} reads here,
\begin{equation}
c^2 k^2 = \omega^2 \mp 2 \Omega_3 \omega.
\end{equation}
From this equation we easily deduce the phase speed 
$v_\varphi :=\omega/k$ and the group speed 
$v_g:=\partial \omega /\partial k$, as well as their
Taylor expansion to third-order for small ratios of the rotation rate 
$\Omega_3$ with respect to the wave frequency $\omega$. 
Introducing $\delta$, the ratio of these two frequencies, 
$\delta := \Omega_3/\omega$, we find
\begin{align}
& \dfrac{v_\varphi}{c} = \dfrac{1}{\sqrt{1 \mp 2\delta}} 
= 1 \pm \delta + \textstyle{\frac{3}{2}} \delta^2 + O( \delta^3) ,  
\notag \\
& \dfrac{v_g}{c} = \dfrac{\sqrt{1 \mp 2\delta}}{1 \mp \delta} 
= 1 + \textstyle{\frac{1}{2}} \delta^2 + O(\delta^3).
\end{align}
Clearly, the right circularly-polarized wave is defined for any value 
of the rotation rate whereas the left circularly-polarized wave only 
exists for a limited range of $\Omega _{3}$, with $\omega /2$ as a 
cut-off frequency.
Note also that a left circularly-polarized wave is accelerated when the
Mooney-Rivlin material is put into rotation and that a right 
circularly-polarized wave is slowed down.

To investigate further nonlinear stress-strain responses, we consider 
two types of incompressible materials belonging to the class of 
`neo-Hookean generalized materials'. 
These are materials whose strain-energy function depends only on the 
first invariant: $W = W(I)$. 
For simplicity, we consider that the solids are not 
prestressed ($\lambda = \mu =1$) prior to the rotation and wave propagation 
although this assumption is not essential.

\subsection{Waves in undeformed Gent materials}

Consider the following strain energy density:
\begin{equation}  \label{Gent}
W_\text{G} = - \dfrac{CJ_m}{2} \ln \left( 1- \dfrac{I-3}{J_m}\right),
\end{equation}
where $C(>0)$ is the infinitesimal shear modulus and $J_m$ is a 
material parameter. 
Gent \cite{Gent} introduced the strain energy function $W_\text{G}
$ to take into account the effect of the finite chain length for the
macromolecular chains composing elastomeric materials 
(see also \cite{HS}).
Hence, the parameter $J_m$ has a physical interpretation: it is the 
constant limiting value for $I-3$, and it reflects the mesoscopic 
finite chain length limiting effect. 
As $J_m \rightarrow \infty$, the limiting effect vanishes and the
strain energy density Eq.~\eqref{Gent} tends to that of a neo-Hookean 
solid (Eq.~\eqref{MR} with $D=0$.)

For the motion considered in this Section, $I = 3 + A^2k^2$ and so, the
limiting chain condition imposes $A^2k^2<J_m$. From the strain energy
density Eq.~\eqref{Gent} we find that the response parameters $\alpha$ 
and $\beta$ defined in Eqs.~\eqref{6} are:
\begin{equation}
\alpha = C\frac{J_{m}}{J_m - A^2 k^2}, \quad \beta = 0.
\end{equation}
It follows from the definition Eq.~\eqref{15} of $Q$, written at 
$\lambda = 1$, that the dispersion equation Eq.~\eqref{dispersion} 
reads, for finite-amplitude circularly-polarized harmonic waves in a 
rotating undeformed Gent material, as
\begin{equation}
C \frac{J_m}{J_m - A^2 k^2} k^2 = \rho (\omega^2 \mp 2\Omega_3 \omega).
\end{equation}
Introducing $\delta := \Omega_3/\omega$, we find that the phase 
velocity $v_\varphi := \omega /k$ is given by
\begin{equation}  \label{GH6}
\rho v_\varphi^2 = 
  \frac{CJ_m + \rho \omega^2 A^2 (1 \mp 2\delta)} {J_m(1 \mp 2\delta)},
\end{equation}
and is defined everywhere for the right wave and only below the cut-off
frequency for the right wave. 
The group velocity, $v_g := \partial \omega /\partial k$, is found as
\begin{equation}
v_g = 
 \frac{ \rho v_\varphi^3}{C} \frac{(1 \mp 2\delta)^2}{1 \mp \delta}.
\end{equation}

In contrast to the case of a Mooney-Rivlin material, the waves are also
dispersive when the body is not rotating; then $\Omega_3 = 0$ and
\begin{equation}
v_\varphi = \sqrt{\dfrac{CJ_m + \rho \omega^2 A^2}{\rho J_m}}, 
\quad 
v_g = \frac{ \rho v_\varphi^3}{C}.
\end{equation}
These latter results are worth mentioning because in \cite{C1}, 
Carroll treated explicitly only the case of Mooney-Rivlin materials. 
Moreover they may be used as benchmarks for an acoustical 
determination of the limiting chain parameter $J_m$. 
Acoustical evaluation is non-invasive and non-destructive, and is
therefore appropriate for an estimation \textit{in vivo}
of $J_m$, whose numerical value can be linked to the ageing and 
stiffening of a soft biological tissue such as an arterial wall 
\cite{HS}.

\subsection{Waves in undeformed power-law materials}

Now consider the following strain energy density,
\begin{equation}  \label{power}
W_\text{K} = 
 \frac{C}{b}\left[ \left( 1+\frac{b}{n}\left( I-3\right) \right)
^{n}-1\right] ,
\end{equation}
where $C(>0)$, $b$, and $n$ are constitutive parameters. 
Knowles \cite{Kn} proposed that this strain energy could account for 
\textit{strain softening} when $n<1$ and for \textit{strain hardening} 
when $n>1$. 
These effects have been observed for many real materials.

Here we find that the dispersion equation Eq.~\eqref{dispersion} is 
given by
\begin{equation}
C\left( 1 + \frac{b}{n} A^2 k^2 \right)^{n-1} k^2 
 = \rho (\omega^2 \mp 2 \Omega_3 \omega).
\end{equation}

Taking $n=2$ in Eq.~\eqref{power} as an example of strain energy for a 
hardening material, we find that the corresponding phase and group 
velocities are given by
\begin{align}
& \rho v_\varphi^2 
= \frac{C}{2 (1 \mp 2\delta)} 
 \left[1 + \sqrt{1+2\frac{\rho\omega^2}{C} bA^2(1 \mp 2\delta)}\right],
  \notag \\
& v_g = \frac{C}{\rho v_\varphi (1 \mp \delta)} 
      \sqrt{1 + 2\frac{\rho \omega^2}{C} bA^2(1 \mp 2\delta)}.
\end{align}

The choice $n = \textstyle{\frac{1}{2}}$ in Eq.~\eqref{power} provides 
an example of strain energy for a softening material. 
As pointed out by Knowles \cite{Kn}, this choice is a borderline value 
for $n$, as the material is elliptic but not uniformly elliptic. 
We compute the corresponding phase speed as
\begin{equation}
\rho v_\varphi^2 = \frac{C}{1 \mp 2 \delta} 
  \sqrt{1 + \left[\frac{\rho \omega^2}{C}bA^2(1 \mp 2\delta) \right]^2}
 - \rho \omega^2 b A^2,
\end{equation}
and we omit to display the group speed because its expression is too
cumbersome.

\section{Motions with sinusoidal time dependence}

In this section we consider finite-amplitude shearing motions with a
sinusoidal time-independence,
\begin{equation}
u(z,t) = \phi (z)\cos (\omega t)+\psi (z)\sin (\omega t), 
\quad 
v(z,t)=\phi (z)\sin (\omega t)-\psi (z)\cos (\omega t), 
 \label{std1}
\end{equation}
which are a subcase of Eqs.~\eqref{field2}. For these solutions we have
\begin{equation}
u_{z}^{2}+v_{z}^{2}=\phi ^{^{\prime }2}+\psi ^{^{\prime }2},
\end{equation}
and so the strain invariants Eq.~\eqref{12} are spatially nonuniform 
and constant in time \cite{C1a}. 
The governing equations Eqs.~\eqref{motion} reduce to
\begin{equation}
(Q\phi')' = \rho (\omega^2 + 2\Omega_3 \omega)\phi,
  \quad 
(Q\psi')' = \rho (\omega^2 + 2\Omega_3 \omega)\psi .
  \label{cm0}
\end{equation}
These equations are consistent at $\Omega _{3}=0$ with those derived 
by Carroll \cite{C2} . 
Following his lead, we reduce them to a problem in central
force motion.

We introduce the functions $\Phi(z)$ and $\Psi(z)$ defined by
\begin{equation}
\Phi := Q \phi', \quad \Psi := Q\psi'.  \label{cm1}
\end{equation}
We assume that these latter equalities are invertible as
\begin{equation}
\phi' = \nu \Phi , \quad \psi' = \nu \Psi ,  \label{cm2}
\end{equation}
where \cite{C2,C3} the generalized shear compliance $\nu $ ($>0$) is a
function of the shear stress $\sigma$, itself given by 
$\sigma^2 = \Phi^2 + \Psi^2$.
For example, in the case of a bi-axially deformed Mooney-Rivlin 
material with strain energy Eq.~\eqref{MR}, $\nu$ is constant: 
$\nu_\text{MR} = 1/(C \lambda^2 + D \lambda)$; 
in the case of an undeformed Gent material with
strain energy Eq.~\eqref{Gent}, we find that $\nu$ is given by 
$\nu_\text{G} 
 = (CJ_m / 2 \sigma^2) (\sqrt{1 + (4 \sigma^2)/(C^2J_m)} - 1)$. 
Note that Carroll \cite{C3} proposed expressions for $\nu$ when the 
strain-energy density is expanded up to sixth-order in the invariants 
($I-3$) and $(II-3)$.

Substitution of Eq.~\eqref{cm2} into the derivative with respect to 
$z$ of Eqs.~\eqref{cm0} leads to the system of coupled ordinary 
differential equations,
\begin{equation}
\Phi'' - \rho (\omega^2 + 2\Omega_3 \omega)\nu \Phi = 0,
  \quad 
\Psi'' - \rho (\omega^2 + 2\Omega_3 \omega )\nu \Psi =0.
  \label{cm3}
\end{equation}
This system is formally equivalent to the one governing the motion of a
particle in a plane under a field of central forces, after 
identification of $\Phi $ and $\Psi $ with the rectangular Cartesian 
coordinates and of $z$ with time. 
The usual change of variables from rectangular Cartesian to polar
coordinates,
\begin{equation}
\Phi =r\cos \theta ,\quad \Psi =r\sin \theta ,  \label{cm4}
\end{equation}
leads to
\begin{equation}
r'' - r \theta^{'2} = 
 \rho (\omega^2 + 2 \Omega_3 \omega) \nu(r^2) r, 
\quad 
r \theta'' + 2 r' \theta' = 0.  \label{cm5}
\end{equation}
These equations coincide at $\Omega_3 = 0$ with those of Carroll 
\cite{C2}.
Eq.~\eqref{cm5}$_2$ is 
integrated as $r^2 \theta' = A$, a constant.
Substituting this new equation into Eq.~\eqref{cm5}$_1$, multiplying 
across by $r'$, and integrating yields
\begin{equation}
r^{' 2} + A r^{-2} 
 - \rho (\omega^2 + 2\Omega_3 \omega )\textstyle{\int} \nu(s)ds 
 = B,
\label{en}
\end{equation}
another constant.
For a further treatment and discussions on the interpretation 
of the solution to this equation, we refer to the papers by 
Carroll \cite{C1a,C2,C3,C4,C5}, at least as long as 
$\Omega_3 > -\omega/2$. 
We note that the nature of this equation and of its solutions is 
dramatically altered as $\Omega_3$ tends to $-\omega/2$ and 
beyond, where it is reasonable to expect that (for example) 
what was a periodic solution to Eq.~\eqref{en} for  
$\Omega_3 > -\omega/2$ has turned into an unbounded solution
for $\Omega_3 < -\omega/2$ because then, the central force of 
Eq.~\eqref{cm5}$_1$ is repulsive instead of attractive.

\section{Motions with sinusoidal spatial dependence}

Finally, we consider a plane wave motion with sinusoidal spatial 
variations,
\begin{equation}
u(z,t) = \phi(t) \cos(kz) + \psi(t) \sin(kz), 
 \quad 
v(z,t) = \phi(t) \sin(kz) - \psi(t) \sin(kz).
\end{equation}
This standing wave \cite{C2,C3} generalizes the superposition of two
circularly-polarized wave propagating in opposite directions. 
It is a subcase of Eqs.~\eqref{field1}.

Here,
\begin{equation}
u_z^2 + v_z^2 = k^2 (\phi^2 + \psi^2),
\end{equation}
so that $I$, $II$, $\alpha$, $\beta$, and $Q$ are independent of $z$. 
The governing equations Eqs.~\eqref{motion} reduce to the system of 
ordinary differential equations,
\begin{equation}
\rho \phi'' + 2 \rho \Omega_3 \psi' + k^2 Q \phi = 0, 
\quad 
\rho \psi'' - 2 \rho \Omega_3 \phi' + k^2 Q \psi =0.  
\label{ss0}
\end{equation}
This system coincides at $\Omega_3 = 0$ with the system 
established by Carroll \cite{C2}. 
It is worth noting that the change of variables,
\begin{equation}
k \phi(z) = r(z) \cos(\theta(z) + \Omega_3 z), 
 \quad 
k \psi(z) = r(z) \sin(\theta(z) + \Omega_3 z),  
 \label{ss1}
\end{equation}
leads to a \emph{modified} central field problem
\begin{equation}
 r'' - r \theta^{'2} + [(k^2/\rho) Q(r^2) + \Omega_3^2] r = 0,
  \quad
 r \theta'' + 2 r' \theta' = 0.  \label{ss4}
\end{equation}

Again, integration of the second equation Eq.~\eqref{ss4}$_2$ 
leads to $r^2 \theta' = A$, a constant. 
Then, substitution into Eq.~\eqref{ss4}$_1$, multiplication, and 
integration leads to 
\begin{equation}
r^{' 2} + A r^{-2} 
 + (k^2/\rho)  \textstyle{\int} Q(s)ds
   + \Omega_3 r^2  
    = B,
\end{equation}
another constant.
Here the presence of rotation $\Omega_3 \neq 0$ always alters the 
nature of the solution with respect to non-rotating case.

\section{Concluding remarks}

Incompressible nonlinear elasticity provided a coherent framework where the
equations of motion could be written in full, and possibly solved, for a rotating 
elastic body, without having to be split into a ``time-dependent'' solution and a 
a hypothetical ``time-independent'' solution.
The internal constraint of incompressibility played a crucial role in the 
writing of these equations, because the arbitrary pressure term can englobe 
the possibly troublesome centrifugal force. 

As an illustration, the equations of motion were solved using 
the finite amplitude motions introduced and developped 
by Carroll in non-rotating elastic bodies.
Because his solutions constitute one of the few examples of finite amplitude
exact solutions, much emphasis was placed on how to derive them.
In particular, it was shown how the search for separable solutions 
could recover and extend Carroll's results. 
For circularly polarized harmonic waves, the dispersion equation was derived 
explicitly and solved for the Mooney-Rivlin, Gent, and power-law strain 
energy functions. 
For motions with sinusoidal time dependence and for motions with 
sinusoidal space dependence, the procedure of reduction to a set of 
ordinary differential equations was outlined. 
Their eventual resolution can be adapted from Carroll's works, 
but is beyond the scope of this contribution.

The resolution of the full equations of motion in a rotating hyperelastic 
\textit{compressible} material is also left open.




\begin{thebibliography}{99}

\bibitem{ScCe73}   
  Schoenberg, M., Censor, D.: 
  Elastic waves in rotating media, 
  Quart. Appl. Math. \textbf{31}, 115--125 (1973).

\bibitem{AhKh01}
  Ahmad, F., Khan, A.: 
  Effect of rotation on wave propagation 
  in a transversely isotropic medium, 
  Math. Probl. Eng. \textbf{7}, 147--154 (2001).

\bibitem{ZhJi01} 
  Zhou, Y.H., Jiang, Q.:
  Effects of Coriolis force and centrifugal force 
  on acoustic waves propagating along the surface of a
  piezoelectric half-space. 
  Z. angew. Math. Phys. \textbf{52}, 950--965 (2001).

\bibitem{Auri04}
  Auriault, J.-L.: 
  Body wave propagation in rotating elastic media, 
  Mech. Res. Comm. \textbf{31}, 21--27 (2004).

\bibitem{Dest04} 
  Destrade, M.: 
  Surface acoustic waves in rotating orthorhombic crystals,
  Proc. Roy. Soc. London \textbf{A460}, 653--665 (2004).

\bibitem{C1} 
  Carroll, M.M.: 
  Some results on finite amplitude elastic waves,
  Acta Mech. \textbf{3}, 167--181 (1967).

\bibitem{C1a} 
  Carroll, M.M.:
  Oscillatory shearing of nonlinearly elastic solids, 
  Z. angew. Math. Phys. \textbf{25}, 83--88 (1974).

\bibitem{C2} 
  Carroll, M.M.: 
  Plane elastic standing waves of finite amplitude, 
  J. Elast. \textbf{7}, 411--424 (1977).

\bibitem{C3} 
  Carroll, M.M.: 
  Reflection and transmission of circularly
  polarized elastic waves of finite amplitude, 
  J. Appl. Mech. \textbf{46}, 867--872 (1979).

\bibitem{C4} 
  Carroll, M.M.: 
  Unsteady homothermal motions of 
  fluids and isotropic solids, 
  Arch. Ration. Mech. An. \textbf{53}, 218--228 (1974).

\bibitem{C5} 
  Carroll, M.M.: 
  Plane circular shearing of incompressible fluids and solids, 
  Q. J. Mech. Appl. Math. \textbf{30}, 223--234 (1977).

\bibitem{C6}
  Curran, J.H., Carroll, M.M.:
  Sinusoidal shearing of liquid crystals, 
  Int. J. Eng. Sci. \textbf{18}, 971--977 (1980).

\bibitem{C7} 
  Carroll, M.M.: 
  On circularly-polarized nonlinear electromagnetic waves, 
  Q. Appl. Math. \textbf{25}, 319--323 (1967).

\bibitem{C8} 
  Carroll, M.M., McCarthy, M.F.: 
  Finite amplitude wave propagation in magnetized perfectly 
  electrically conducting elastic materials, 
  In: (McCarthy, M.F., Hayes, M.A., eds), 
  Elastic Wave Propagation, North-Holland: Elsevier 1989, pp. 615--621.

\bibitem{RR} 
  Rosenau, P.,  Rubin, M.B.: 
  Motion of a nonlinear string: some exact solutions to an old 
  problem, 
  Phys. Rev. A \textbf{31}, 3480--3482 (1985).

\bibitem{BB} 
  Beatty, M.F.:
  Topics in finite elasticity: Hyperelasticity of rubber, elastomers, 
  and biological tissues - with examples, 
  Appl. Mech. Rev. \textbf{40}, 1699--1733 (1987). 

\bibitem{Liu02}  
  Liu, I.-S.:
  Continuum Mechanics, 
  Wien New York : Springer 2002.

\bibitem{O} 
  Olver, P.J.: 
  Applications of Lie Groups to Differential Equations, 
  2nd ed. Wien New York: Springer 1993.

\bibitem{V} 
  Vassiliou, P.: 
  Coupled systems of nonlinear wave equations and
  finite dimensional Lie Algebras I, II 
  Acta Appl. Math. \textbf{8}, 107--163 (1987).

\bibitem{BoHa95}
  Boulanger, Ph., Hayes, M.: 
  Further properties of finite-amplitude plane waves 
  in deformed Mooney-Rivlin materials, 
  Quart. J. Mech. Appl. Math. \textbf{48}, 427--464 (1995).

\bibitem{Gent}
  Gent, A.:
  A new constitutive relation for rubber, 
  Rubber Chem. Technol. \textbf{69}, 59--61 (1999).

\bibitem{HS} 
  Horgan, C.O., Saccomandi, G.: 
  A description of arterial wall mechanics 
  using limiting chain extensibility constitutive models, 
  Biomechan. Model. Mechanobiol. \textbf{1}, 251--266 (2003).

\bibitem{Kn} 
  Knowles,  J.: 
  The finite anti-plane shear field near the tip of a crack for a 
  class of incompressible elastic solids, 
  Int. J. Fracture \textbf{13}, 611--639 (1977).

\end{thebibliography}
\end{document}